# Capillary Filling of Polymer Chains in Nanopores

*Jianwei Zhang [1], Jinyu Lei [1], Wenzhang Tian [1], Guangzhao Zhang [1], George Floudas [2,3,4], and Jiajia Zhou [5,6,*]*

1. Faculty of Materials Science and Engineering, South China University of Technology, Guangzhou 510640, China

2. Max Planck Institute for Polymer Research, 55128 Mainz, Germany

3. Department of Physics, University of Ioannina, 45110 Ioannina, Greece

4. Institute of Materials Science and Computing, University Research Center of Ioannina (URCI), 45110 Ioannina, Greece

5. South China Advanced Institute for Soft Matter Science and Technology, School of Emergent Soft Matter, South China University of Technology, Guangzhou 510640, China

6. Guangdong Provincial Key Laboratory of Functional and Intelligent Hybrid Materials and Devices, South China University of Technology, Guangzhou 510640, China

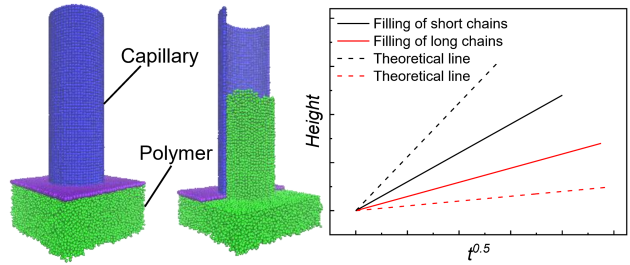

For Table of Contents use only.

**ABSTACT:** We performed molecular dynamics simulations with a coarse-grained model to investigate the capillary filling dynamics of polymer chains in nanopores. Short chains fill slower than predicted by the Lucas-Washburn equation but long chains fill faster. The analysis shows that the combination of the confinement effect on the free energy of chains and the reduction of the effective radius due to the "dead zone" slow down the imbibition. Reduction of the entanglements is the main factor behind the reversing dynamics because of the lower effective viscosity, which leads to a faster filling. This effect is enhanced in the smaller capillary and more profound for longer chains. The observed increase in the mean square radius of gyration during capillary filling provides a clear evidence of chain orientation, which leads to the decrease in the number of entanglements. For the scaling relation between the effective viscosity and the degree of polymerization, we find the exponent will increase in the larger nanopore.

**INTRODUCTION**

A better understanding of the polymer chain dynamics in nanopores is of profound significance for protein ultrafiltration[1], modern lab-on-chip applications[2], and enhancement of the mechanical strength in nanocomposite materials[3,4]. One hundred years ago, Lucas and Washburn derived the celebrated equation (LWE) to describe the imbibition dynamics of viscous Newtonian fluids in a cylindrical capillary,[5,6]

$$h(t) = \sqrt{\frac{\gamma R \cos\theta}{2\eta} t} \qquad (1)$$

where $h(t)$ is the filling length of fluid inside the pore as a function of wetting time $t$, $\gamma$ is the surface tension of fluid, $R$ is the capillary radius, $\theta$ is the equilibrium contact angle of fluid on the capillary wall, and $\eta$ is the bulk fluid viscosity. The LWE was verified by several experiments at macroscopic scales.

An implicit but indispensable assumption in the derivation of the LWE is that the capillary radius should be an order of magnitude larger than the size of fluid molecules. Only under this condition the fluid can be treated as a continuum medium.[6] The breakdown of LWE in channels of nanometer scale is then well expected and understandable. Experimentally, deviation from the theoretical prediction of LWE in the case of polymer melts confined in nanotubes has been observed in several studies.[7,8] Provided that spatial constraints are imposed on polymer chains, experiments exhibit non-classical behavior with respect to mobility[9], entanglement[10], viscosity[11], etc. Henceforth, the LWE might not be applicable at the nanoscale. Interestingly, the scaling relationship between $h$ and $t^{1/2}$ still persists. However, the prefactor $\sqrt{\gamma R \cos\theta / 2\eta}$ (marked as $A$ in rest of the paper) is different from the LWE prediction.

According to recent experiments, the chain length has a profound effect on the difference between the experimental value and theoretical value of the prefactor. Short polymer chains, such as poly(ethylene oxide) with molar mass below 100 kg/mol, shows a slower imbibition than the theoretical prediction in self-ordered nanoporous aluminum oxide (AAO) with radius $R \approx 10 - 100$ nm.[7] Nevertheless, reversal in the imbibition dynamics occurs for longer polymer chains within the same nanopores. In the latter case, capillary filling is faster than theoretically predicted. The phenomenon is not only observed inside capillaries of regular shape, but also shown in the irregular shaped nano-channels in dense packings of nanoparticles by Lee *et al*.[12-14] This is deeply associated with the application of polymer nanomaterials with fillers.[15,16] The non-monotonic dependence of the Lucas-Washburn prefactor on the chain length was also confirmed by in situ monitoring of the imbibition using dielectric spectroscopy.[8,17-19] However, the physics behind this phenomenon is still under investigation. One feasible theoretical explanation proposed recently[20] suggested that it is the combination of the confinement effect on the free energy of chains and the reduction of the effective radius due to the "dead zone". The dead zone is formed by the strong adsorption of polymer chains on the capillary wall, which triggers slower capillary filling. Faster imbibition occurs when polymer chains are well entangled. Chains are constrained by other chains; they can only move along the "reptation tube". Under confinement, the chains are less entangled which leads to a smaller effective viscosity. Therefore, the filling will be accelerated.

Several molecular dynamics simulation models were used to explore the microscopic origin of this phenomenon. Vo *et al*.[21] attributed the slower uptake of the fluid than the theoretical predictions to liquid layering in the vicinity of the solid surface. The overall effect results in a higher density and a larger viscosity. Shavit *et al*.[22] also found an increased viscosity near the supporting substrate by calculating the local viscosity of polymers. Dimitrov *et al*.[23] demonstrated that the existence of slippage accelerates the imbibition and suggested a simple modification of the LWE so as to work quantitatively even at the nanoscale. Henrich *et al*.[24] discussed the importance of the dynamic contact angle and precursor film in slits of width within 2 μm. Several dissipative particle dynamics models also suggested that the above effects should be considered to improve LWE.[25,26] However, the length of polymer chains is another significant factor that has been overlooked in capillary filling studies. Ring *et al*.[27] have investigated the effect of the length of polymer chains on the critical contact angle to imbibition. However, a systematic exploration of the chain-length effects on the reversal of capillary filling is absent. The aim of the present work is to verify the mechanism of dynamics reversal and to identify the underlying microscopic evidence.

In this work, we employ extensive molecular dynamics simulations on the capillary filling of a polymer melt in a nanopore and systematically investigate the chain-length effects on the prefactors of LWE. The article is organized as follows: The Simulation Model section collects the basic polymer properties required by LWE and calculates the theoretically predicted value. The Results and Discussion section contrasts the simulation results of polymer melt imbibing in a nanopore against the LWE predictions. Finally, we present the analysis of chain-length effects on the capillary filling dynamics and conclude with a summary.

**SIMULATION MODEL**

We performed coarse-grained molecular dynamics simulations to study the imbibition dynamics. The capillary wall is modeled as ordered spherical beads and the polymer chains are presented by a

bead–spring model. Non-bonded interactions between all beads were modeled by the truncated-shifted Lennard-Jones (LJ) potential[28]

$$U_{LJ}(r) = \begin{cases} 4\varepsilon_{LJ}\left[\left(\frac{\sigma}{r_{ij}}\right)^{12} - \left(\frac{\sigma}{r_{ij}}\right)^{6} - \left(\frac{\sigma}{r_{cut}}\right)^{12} + \left(\frac{\sigma}{r_{cut}}\right)^{6}\right] & r_{ij} \leq r_{cut} \\ 0 & r_{ij} \geq r_{cut} \end{cases} \quad (2)$$

where $\sigma$ is the bead diameter, $\varepsilon_{LJ}$ is the Lennard-Jones interaction parameter, $r_{ij}$ is the distance between the $i$th and $j$th beads, and the cutoff distance $r_{cut}$ was set to $2.5\,\sigma$. For interactions between polymer beads, $\varepsilon_{LJ} = 1.4\,k_B T$ and $\sigma = 1.0$, where $k_B$ is the Boltzmann constant and $T$ is the absolute temperature. For interactions between beads of wall atoms, in order to prevent polymer chains from penetrating the wall and to add a moderate spring force to fix wall beads, we set $\varepsilon_{LJ} = 1.0\,k_B T$ and $\sigma = 0.8$.[23] The bonds connecting beads into polymer chains were modeled by the finite extension nonlinear elastic (FENE) potential[29]

$$U_{FENE}(r) = -\frac{1}{2}k_{spring}R_{max}^{2}\ln\left(1 - \frac{r^2}{R_{max}^{2}}\right) \quad (3)$$

with the spring constant $k_{spring} = 30\,k_B T/\sigma^2$ and the maximum bond length $R_{max} = 1.5\,\sigma$. The repulsive part of the bond potential was modeled by the LJ potential with $r_{cut} = 2.5\,\sigma$ and $\varepsilon_{LJ} = 1.4\,k_B T$. Macromolecules with degrees of polymerization of $N = 1, 10, 50, 100, 200, 280, 400$ were studied in our simulation.

Cylindrical nanopores with different capillary radii ($R = 5\,\sigma,\ 10\,\sigma,\ \text{and}\ 15\,\sigma$) were modeled by one layer of wall atoms respectively. Their length was set to $L = 70\,\sigma$. The distance between two neighbor beads of wall atoms was $1.08\,\sigma$. The wall atoms may vibrate around their initial lattice position under a spring force $F = -Kr$, where $r$ is the displacement of the atom from its current position to its initial position, and $K$ is a constant that we set to $300\ k_BT/\sigma^2$. Under the capillary reside a reservoir of polymer melt with periodic boundaries perpendicular to the tube axis (shown in Fig. 1). The total number of polymer beads is 28000 in the cases of $R = 5\ \sigma$ and $R = 10\ \sigma$, and 40000 in the case of $R = 15\ \sigma$. Initially, the reservoir was sealed to prevent polymers from moving into the capillary. Following the polymer melt equilibration, we deleted beads in the center area to connect the capillary to the reservoir, and the imbibition starts. The height of the capillary filling of the polymer melt was calculated by fitting the density profile to the function

$$\rho(h) = \frac{\rho_p + \rho_v}{2} - \frac{\rho_p + \rho_v}{2} \tanh\left[\frac{2(h - h_0)}{l}\right] \qquad (4)$$

as shown in Fig. 2, where $\rho_p$ is the density of polymer melt in the capillary, $\rho_v \approx 0$ is the vapor density in the capillary due to the low polymer vapor pressure, $h_0$ is the height of interface between polymer melt and its vapor, and $l$ is the thickness of the interface. The relationship between height $h$ and time $t$ of polymer melt in the capillary, with the interaction parameter between chains and wall atoms $\varepsilon_{LJ} = 1.6\ k_BT$ and $\sigma = 1.0$, was then monitored.

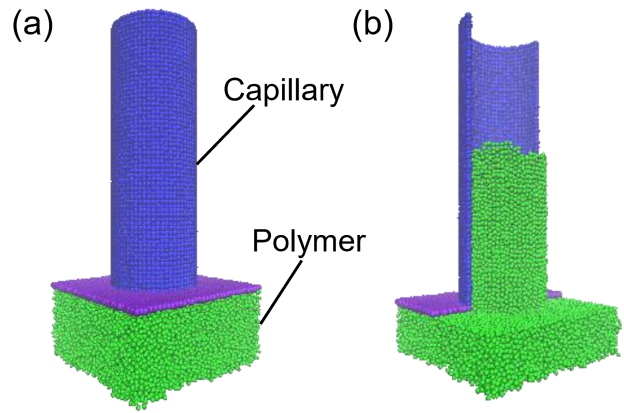

**Figure 1.** Snapshots of capillary imbibition model. (a) Initial state of capillary imbibition. (b) Capillary imbibition in progress (half of the tube was cut off for better view).

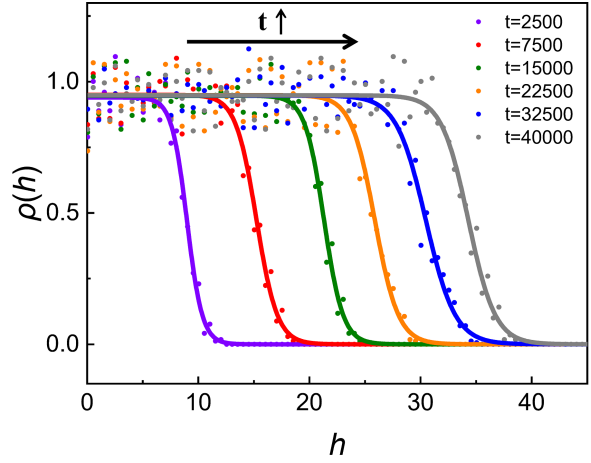

**Figure 2.** Profiles of the average polymer melt density $\rho(h)$ during capillary filling in the nanopore with $R = 10\,\sigma$ at various times in the case of $N = 400$.

To contrast with theoretical relationship $h(t)$ of Lucas-Washburn equation, we calculated in addition, the surface tension $\gamma$, the viscosity $\eta$, and the contact angle $\theta$ of the polymer melt as function of the chain length $N$.

**Surface tension.** We employed the Irving-Kirkwood's expression[30] to calculate the surface tension of the polymer melt. All data were derived from a flat gas-liquid interface by integrating the asymmetric part of the pressure tensor.

$$\gamma = \int_{-\lambda}^{\lambda} [P_N(z) - P_T(z)]\, dz \tag{5}$$

Here, $2\lambda$ is the thickness of whole gas-liquid interface determined by the variation of $P_N(z) - P_T(z)$, $P_N(z)$ is the normal component and $P_T(z)$ is the tangential component to the interface of the pressure tensor. This integrand should be zero both in gas phase and bulk phase, as shown in Fig. S1.

**Viscosity.** The density of the bulk liquid was evaluated from the density profile of the reservoir before imbibition in Fig. S2 (the result is presented in Table S1). The viscosity of the polymer melt was calculated by creating a Poiseuille flow.[31] It was obtained by setting up two plates being parallel to $xy$-plate. Subsequently, a polymer melt with density shown in Table S1 was filled between plates. Then a body force along $x$-axis was applied on every monomer as shown in Fig. S3. Following the steady state, we fit

$$v_x(z) = \frac{\rho g_x}{2\eta}(C + Dz - z^2) \qquad (6)$$

to the velocity profile along $z$-axis in Fig. S4, where $v_x$ is the velocity being parallel to the force, $\rho$ is the density, $g_x$ is the body force acting on the polymer, $\eta$ is the viscosity of the polymer melt, $C$ is a constant equal to zero when there is no-slip boundary condition, and $D$ is the distance between two plates. Different values of $g_x$ in the range from $0.2\ k_BT/\sigma$ to $0.8\ k_BT/\sigma$ were employed to validate the result.

Another method to obtain the viscosity is from the Green-Kubo (GK) integral formula[32-34]

$$\eta = \frac{V}{k_BT}\int_0^\infty \langle \tau_{\alpha\beta}(t_0)\tau_{\alpha\beta}(t)\rangle dt \qquad (7)$$

where $V$ is the volume and $\tau_{\alpha\beta}$ stands for an off-diagonal element of the stress tensor. The calculations under equilibrium conditions were made in a $30\times30\times30$ box with periodic boundaries, which was filled with polymers at the same densities with the former method. The viscosity values obtained from Green-Kubo formula are shown in Fig. S5 and compared with the results from Poiseuille flow. Despite Green-Kubo calculation is with large margin of error[31], the results of the two approaches agree well.

**Contact angle.** The contact angle of the polymer melt on the capillary wall was calculated by putting the same polymer droplet on a plate that has the same lattice structure and interaction parameters with the capillary. In this part, we set the interaction parameter between the polymer and wall atoms as $\varepsilon_{LJ} = 1.2, 1.4, 1.6, 1.8\ k_B T$ and $\sigma = 1.0$. After the droplet reaches equilibrium, we fitted the shape of the droplet in Fig. S6 and a series of contact angles were obtained.

All simulations were performed using LAMMPS[35] and carried out under a constant temperature maintained by DPD thermostat[36] with friction parameter $\xi = 0.5\ \sqrt{mk_B T/\sigma^2}$ and interaction cutoff $r_{cut} = 2.5\ \sigma$. The friction parameter $\xi$, determines the strength of the dissipative force and reflects the efficiency of keeping the temperature constant following thermal disturbance. We used the velocity-Verlet algorithm[37] with a time step $\Delta t = 0.005\ \tau_{LJ}$ for integration, where $\tau_{LJ} = \sqrt{m\sigma^2/\varepsilon_{LJ}}$ is the standard LJ-time.

**RESULTS AND DISCUSSION**

We show the simulation results of the surface tension $\gamma$, viscosity $\eta$, and contact angle $\theta$ of the polymer in Fig. 3, Fig. 4, and Fig. 5, respectively. The variation of surface tension $\gamma$ with molecular weight $M$ for polymer melt follows a relation $|\gamma - \gamma_\infty| \propto M^{-1}$, which has been confirmed by many experiments[38], simulations[39], and theory[40,41]. Our result is consistent with this relation as shown in Fig. 3(b), where $N = M/M_0$ and $M_0$ is the molar mass of the repeat unit.

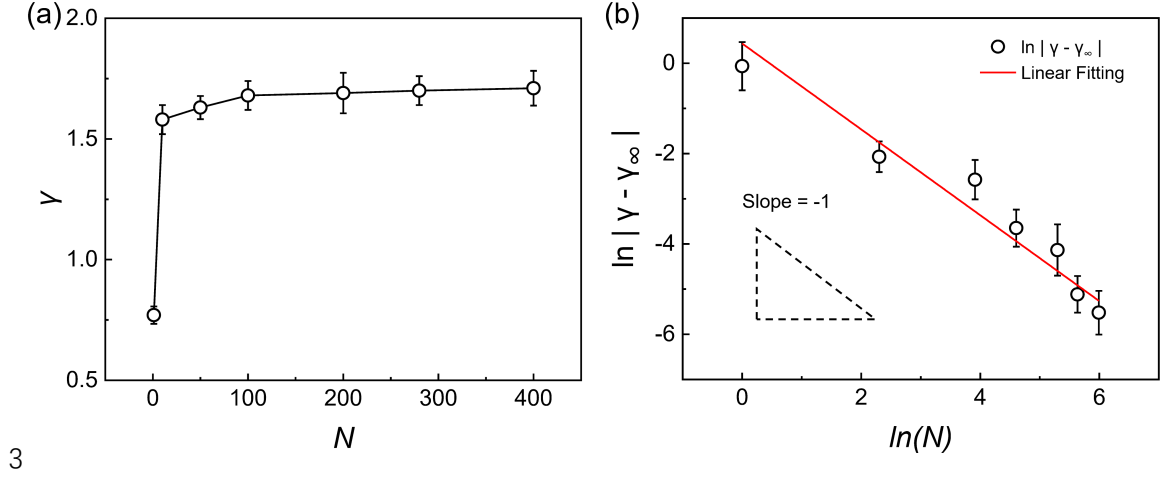

**Figure 3.** (a) Surface tension, $\gamma$, of a polymer melt as a function of the degree of polymerization, $N$. (b) Result of linear fitting of the data after taking the logarithm. The slope of the line, representing the exponent of the scaling relationship, is approximately $-1$.

The viscosity of the polymer melt is also a function of molar mass.[42] It is widely accepted that the relationship can be described as $\eta \propto M^{1.0}$ ($M < M_c$) at relatively low molar mass and $\eta \propto M^{3.4}$ ($M > M_c$) at very high molar mass, where $M_c$ is the critical molar mass. The rapid increase in viscosity above $M = M_c$ is explained by the increasing number of entangled chains.[43] In simulation, we find a good linear fitting when $N$ is relatively low, as shown in Fig. 4. For higher $N$, entanglements increase the viscosity. According to previous works, the entanglement length $N_e$ of flexible-chain systems similar with our model, is about 84,[44] and the steep increase in viscosity only appears when $N > 5N_e$.[45] Therefore, the longest-chain system ($N = 400$) in our simulation has a moderate level of entanglements.

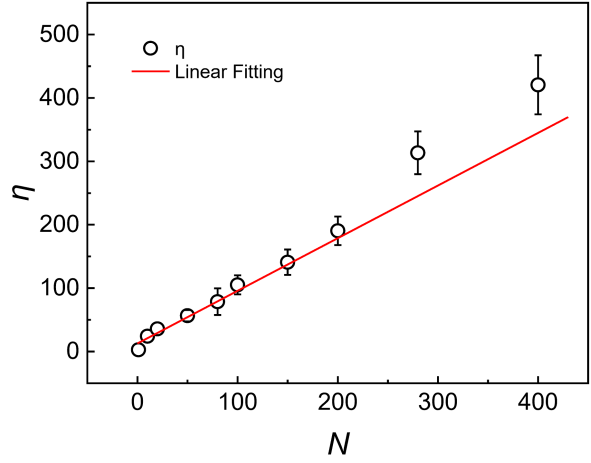

**Figure 4.** The viscosity $\eta$ of polymer melt as a function of the degree of polymerization, $N$.

With respect to the contact angle, $\theta$, we obtain a linear relationship between $\cos\theta$ and the solid-liquid interaction parameter $\varepsilon_{LJ}$ in the case of partial wetting, as shown in Fig. 5(a). For a given value of $\varepsilon_{LJ}$ ($\varepsilon_{LJ} = 1.6$ here), $\theta$ increases and $\cos\theta$ decreases with increasing $N$ as shown in Fig. 5(b). Fitting the expression for the $N$ dependence of the contact angle[27] $1 + \cos\theta = W_{ad}/[\gamma_1 + b\rho_0 k_B T \left(\Gamma_\infty - \frac{2A}{N}\right)]$ to the data, where $W_{ad}$ is the work of adhesion between the polymer and surface, $\gamma_1$ is the component of surface tension except entropic part and is about zero, $b$ is the Kuhn length, $\rho_0$ is the monomer density, $\Gamma_\infty$ is the surface tension of infinitely long polymers, and $A$ is the contribution to surface tension from a single chain end, we obtain the work of adhesion $W_{ad} = 0.353\ k_B T$ at $\varepsilon_{LJ} = 1.6\ k_B T$.

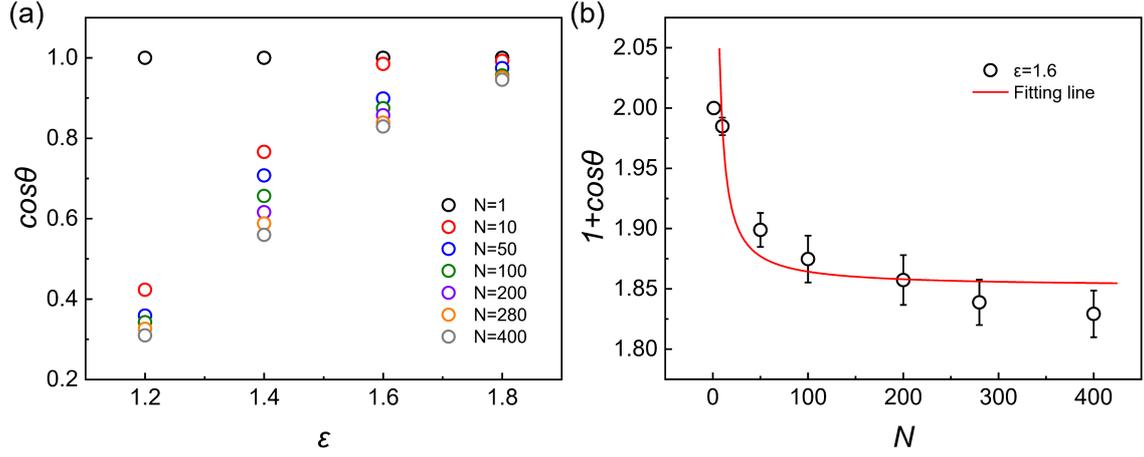

**Figure 5.** (a) Contact angle $\theta$ of the polymer melt as a function of $N$ at $\varepsilon_{LJ} = 1.2, 1.4, 1.6, 1.8\ k_BT$ respectively. (b) Relationship between contact angle $\theta$ and the degree of polymerization $N$. The red line is the result of fitting at $\varepsilon_{LJ} = 1.6\ k_BT$ by using the equation: $1 + \cos\theta = W_{ad}/[\gamma_1 + b\rho_0 k_B T \left(\Gamma_\infty - \frac{2A}{N}\right)]$.

So far, we collected all parameters required by LWE. Next, we proceed to calculate the theoretically predicted relationships of $h(t)$, shown as dashed lines in Fig. 6. The results of the actual imbibition height obtained from simulations are plotted and linked by solid lines in Fig. 6. A non-monotonic change of deviation from the theoretical lines with increasing chain length can be observed. In the case of short chains, the imbibition process is slower than the theoretical prediction, as the calculated points are below the dash lines. In the case of long chains, filling is faster than the theoretically prediction, as the simulation points are now above the dashed lines.

To make the trend clearer, we plot the ratios of actual prefactor $A_a$ and theoretical prefactor $A_t$ at different $N$ in Fig. 7(a). When $N = 1$, there is no bond interaction between the beads and the fluid behaves like a Newtonian fluid. As a result, the imbibition at $N = 1$ agrees with the LWE prediction in all three cases. For polymers $N > 1$, and in the case of the strongest confinement, e.g. within the smallest nanopore ($R = 5\ \sigma$) the reversal in dynamics (defined as the transition from $A_a < A_t$ to $A_a > A_t$) of capillary filling occurs at small chain lengths (between $N = 50$ and $N = 100$). In the nanopore with the intermediate size ($R = 10\ \sigma$), there is a slower penetration velocity for small $N$ up to $N \approx 100$. The reversal in dynamics of capillary filling is found when $N \approx 200$. The capillary filling is then faster than theoretically predicted and $A_a$ is higher than $A_t$. In the nanopore with the largest radius ($R = 15\ \sigma$), the reversal in the imbibition dynamics could not be observed even for $N \approx 400$. For $N < 400$, the actual imbibition is slower than theoretically predicted. We also investigate the relation between $A_a/A_t$ and the degree of confinement as shown in Fig. 7(b). The degree of confinement is represented by the ratio of the radius of gyration $R_g$ of chains in the bulk to the capillary radius $R$. The data of $R_g$ are shown in Fig. 11 below. When $R = 5\ \sigma$, as shown as the black curve in Fig. 7(b), long polymer chains suffer a high level of confinement as $R_g \gg R$, and also show an extreme faster-than-LWE filling. When $R = 10\ \sigma$, most chains have $R_g \ll R$ and some long chains have $R_g \sim R$. The dynamics reversal occurs when chains are relatively long. When $R = 15\ \sigma$, polymers in all chain lengths have $R_g \ll R$, and even long-chain systems show a slower-than-LWE filling. In other words, the interesting observation is that the imbibition dynamics reversal seems to occur at a higher value of $R_g/R$ for the smaller pores.

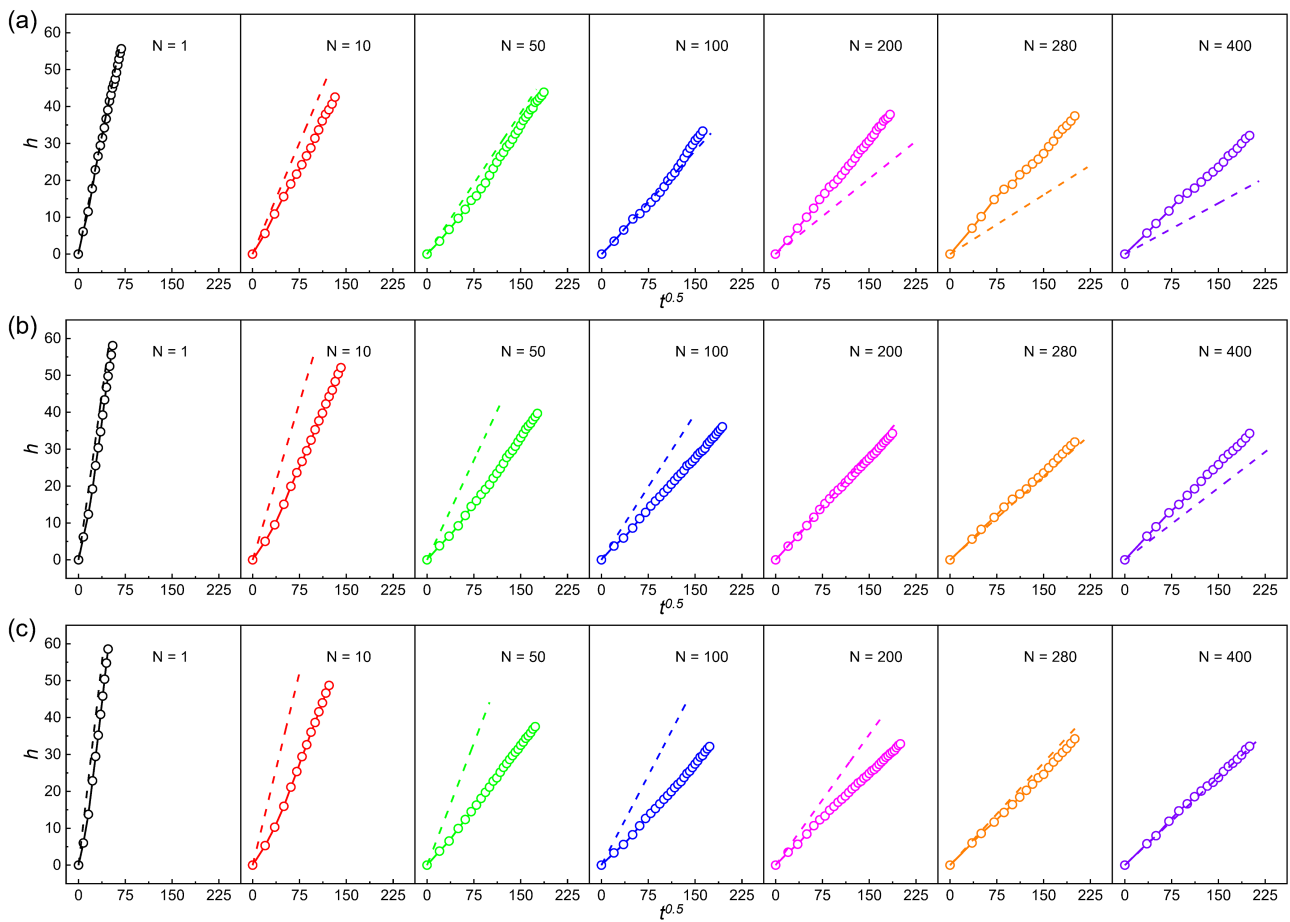

**Figure 6.** Comparison of actual polymer melt capillary filling in nanopores with different radii and LWE theoretical prediction in the cases of $N = 1$ (black lines), $N = 10$ (red lines), $N = 50$ (green lines), $N = 100$ (blue lines), $N = 200$ (magenta lines), $N = 280$ (yellow lines), and $N = 400$ (purple lines), when (a) $R = 5\,\sigma$, (b) $R = 10\,\sigma$, and (c) $R = 15\,\sigma$.

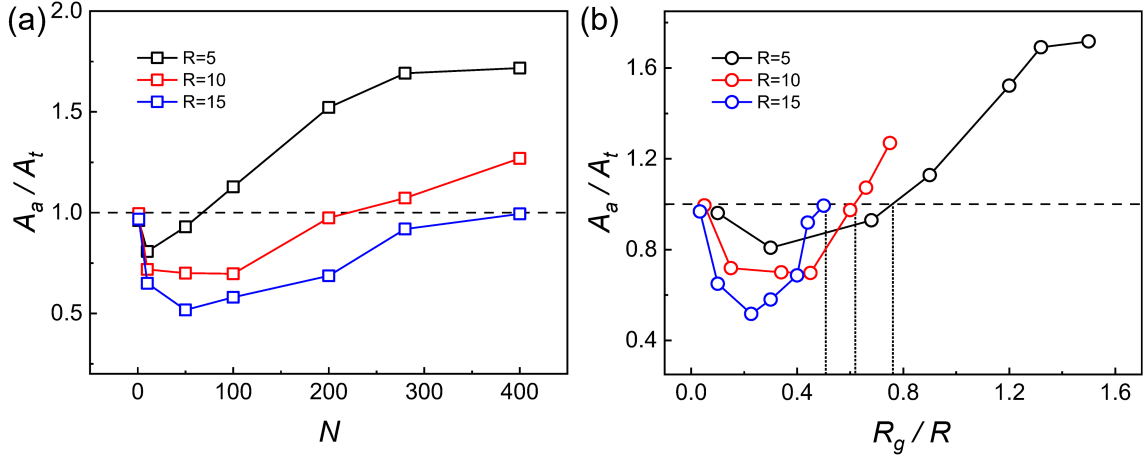

**Figure 7.** (a) Variation of the ratio of actual prefactor $A_a$ and theoretical prefactor $A_t$ with $N$ of all three capillary filling with $R = 5\,\sigma$ (black line), $R = 10\,\sigma$ (red line), and $R = 15\,\sigma$ (blue line). (b) Variation of the ratio of actual prefactor $A_a$ and theoretical prefactor $A_t$ with the degree of confinement (defined as the ratio of the radius of gyration of chains in the bulk $R_g$ and capillary radius $R$) for capillary radii of $R = 5\,\sigma$ (black line), $R = 10\,\sigma$ (red line), and $R = 15\,\sigma$ (blue line).

As wetting requires a strong interaction between the liquid and the solid, there exists a layer of polymer chains attached to the capillary wall, leading to an effective smaller pore radius. This is the dominant effect causing the slower imbibition. On the macroscopic level, the capillary radius is extremely larger than the size of polymer chains, thus the thickness of the adsorption layer can be neglected. Nevertheless, the effect of this layer is significant in nanopores. We explore this factor by collecting the relationship between the instant velocity $V_z$ of beads and the distance $r$ from the center axis of the capillary. To obtain a better average, we create another simulation with new capillaries ($R = 5\,\sigma$, $R = 10\,\sigma$ and $R = 15\,\sigma$) that two ends are periodic boundaries and all other parameters remain the same as before. It is filled with the same chains as above. Additional pressure $\Delta p = 2\gamma/R$ to drive the capillarity is then applied to the polymer melt. To reduce the variables, we take $\Delta p = 0.15, 0.08, 0.05\ k_BT/\sigma^3$ for capillaries of $R = 5, 10, 15\ \sigma$, respectively. Though this method may not fully reproduce the capillary filling quantitatively, we expect qualitative analysis of the effects is sufficient. The relation between $V_z$ and $r$ in the case of $N \leq 100$ is shown in Fig. 8.

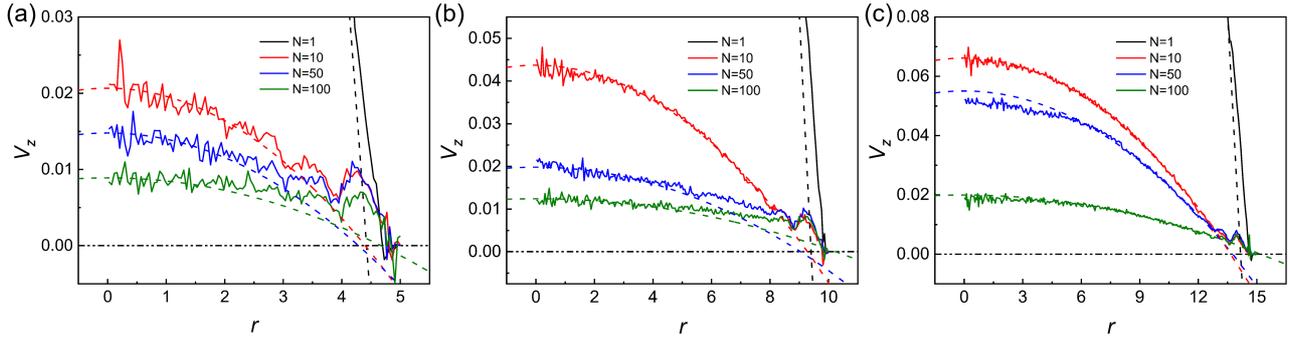

**Figure 8.** Variation of the polymer velocity in $z$ direction, $V_z$, with the distance $r$ from the center of capillary in the new model with periodic boundary. The data are collected in the cases of $N \leq 100$ and (a) $R = 5\ \sigma$, (b) $R = 10\ \sigma$, and (c) $R = 15\ \sigma$. The dash line is the result of fitting by the Poiseuille's formula.

We fit the near-center portion of the velocity profiles to the Poiseuille's formula $V_z = c(-r^2 + R_{eff}^2)$, where $R_{eff}$ is the effective capillary radius. On the macroscopic level, $R_{eff}$ should be equal to the capillary radius $R$. However, it is no longer true at the nanoscale due to the dead zone effect. The velocity profiles in Fig. 8 show deviation from the parabolic shape close to the wall. Note the small bump close to the wall is due to the density variation of polymer beads. For polymeric liquid, Figure 8 indicates that polymers in the dead zone might not be exactly immobile, but loosely adsorbed on the wall with reduced velocity. We show the results of $R - R_{eff}$ as a function of chain length in Fig. 9. In the cases of short-chain systems ($N \leq 50$), we see the value of $R - R_{eff}$ is about 1 to $2\sigma$, and the change in $R - R_{eff}$ can qualitatively give the information of the dead zone. In these cases, the value of $R - R_{eff}$ shows slightly increase with the chain length $N$, suggesting that the dead zone is influenced by the polymer size. This dead zone effect is also shown in video 1 provided in the Supporting Information section. The video shows two chains entering the tube at the same time. However, the chain in the center moves far as compared to the chain near the wall. The latter shows little displacement due to strong adsorption. Video 2 demonstrates the brush-like chains near the wall in the capillary filling. For long-chain systems, we observed some slippage on the wall, which leads to a negative value of $R - R_{eff}$. We attribute this to the difference in the driving force for the flow, as here the flow is driven by the pressure difference, which might differ from the capillary filling system driven by the interfacial forces.

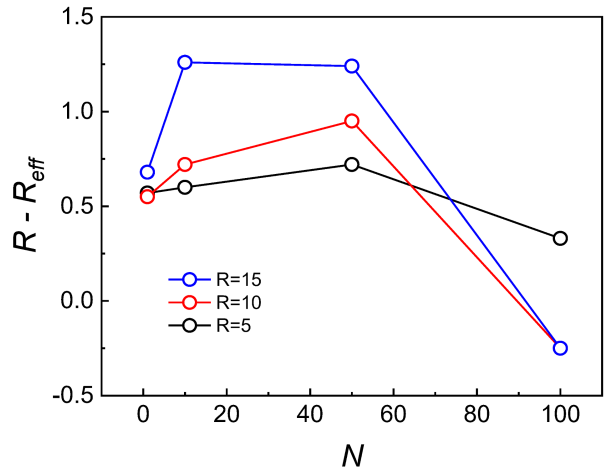

**Figure 9.** Variation of the difference between the nanopore radius and the effective capillary radius $R - R_{eff}$ with the degree of polymerization $N$ at $N \leq 100$.

In general, viscosity determines the kinetic speed of the fluid and entanglements lead to higher viscosity. Therefore, to determine the acceleration effects, we calculate the average number of entanglement points $Z$ per chain in the capillary filling processes for different degrees of polymerization. To this end we follow the primitive path analysis algorithm proposed by Everaers *et al*.[46,47] The specific operation is provided in the Supporting Information. We first collect $Z$ of the chains at $t = 0$, which represents the instance that the chains are in bulk (not yet entering the capillary). We also collect $Z$ of the chains in the Poiseuille flow in the viscosity calculation section to verify whether chains are unentangled when we calculating their viscosity. Additional data are collected after capillary filling reaches a height twice the radius of gyration $R_g$ of chains, at $t = 25000, 30000, 35000, 40000$ respectively. To eliminate the influence of the gas-liquid interface as well as of the liquid not yet entered, we employ the middle part of the flow ($h = 10{\sim}20$) in the capillary to perform the calculation. Figure 10 illustrates the relationship between $Z$ and wetting time $t$ of chains. Combining previous work[44] and our results, we confirm that chains with $N \leq 50$ are free of entanglement, thus not shown in Fig. 10. For $N \geq 100$, $Z$ of the chains during capillary filling decreases in all three cases, as compared to both in the bulk and in the Poiseuille flow in the viscosity calculation section. This effect is more pronounced for higher $N$. Furthermore, the larger reduction in $Z$ is seen when the capillary radius is smaller. This indicates that confinement is a primary reason for the chain disentanglement during flow. Both facts considered, i.e., capillary filling acceleration when $N$ exceeds the requirements of mild entanglement and decrease of $Z$ when chains penetrate the capillary, prove that disentanglement is the main reason for the dynamical reversal. We deem that the chains are oriented by capillary force when filling in nanopores. As a result, the entanglements of longer chains are unwound. According to the works by Yao *et al*.[20] and Venkatesh *et al*.[16], chain transport under very strong confinement is mainly driven by the pressure gradient, where interfacial effects dominate. The strong surface-polymer friction may reduce the number of possible configurations and chains can only move along the "reptation tube". These effects could reduce the number of entanglements and lead to the faster filling speed. Also, researches of multiple-quantum NMR technique experiments[48] and molecular dynamics simulations[10] give the further support. The orientation effect results in a lower fluid viscosity due to fewer entanglements that reduces the resistance to flow. However, this acceleration effect is absent in the case of short chains.

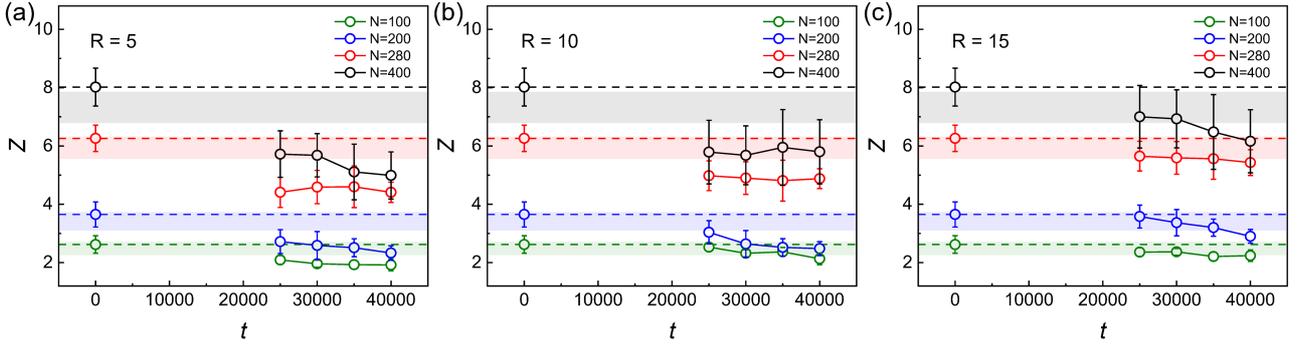

**Figure 10.** Variation of the number of entanglement points, $Z$, with wetting time $t$ during filling in the nanopore with (a) $R = 5\,\sigma$, (b) $R = 10\,\sigma$, and (c) $R = 15\,\sigma$, in the cases of $N = 100$ (green dots), $N = 200$ (blue dots), $N = 280$ (red dots), and $N = 400$ (black dots). The dash lines represent $Z$ of chains in the bulk located outside the capillary. The shaded areas represent $Z$ of chains in the Poiseuille flow in the viscosity calculation section.

The variation of the radius of gyration $R_g$ is another evidence of the chain orientation. We first collect $R_g$ of polymer chains in equilibrium with

$$R_g^2 = \sum_i [(x_i - x_{cm})^2 + (y_i - y_{cm})^2 + (z_i - z_{cm})^2] \qquad (8)$$

where the subscript $i$ stands for the $i$th bead and $cm$ stands for the center of mass of chains. The results are shown in Fig. 11(a). We take the logarithm of our data in Fig. S8 and the slope of the linear fitting is about 0.5. This is consistent with the scaling relationship $R_g \propto N^{0.5}$ of polymer melts.[49] Then, to contrast with equilibrium, we calculate $R_g$ of chains in the capillary during $t = 35000 \sim 40000$, the last 5000 time units in imbibition simulation. The imbibition heights of all groups are now more than twice the radius of gyration $R_g$ of chains in static status, and the average $R_g$ in dynamic processes is shown in Fig. 11(a). It is noticeable that $R_g$ of polymer chains in all three capillaries is larger than in the bulk, which illustrates that the chain orientation takes place during the filling processes. Additionally, the longer the chain, the more prominent the effect of chain orientation. Furthermore, we can see that the effect is more dramatic for chains in the capillary with smaller radius. This means that the confinement is also stronger in the case of smaller nanopore. Components of $R_g$ along the direction of capillary $R_{g\perp}$ and in the radial direction $R_{g\parallel}$ are also collected by separating the tensor of $R_g$ in dynamic processes,

$$R_{g\perp}^2 = \sum_i (z_i - z_{cm})^2 , R_{g\parallel}^2 = \sum_i \frac{1}{2}[(x_i - x_{cm})^2 + (y_i - y_{cm})^2] \tag{9}$$

as shown in Fig. 11(b)(c)(d). In all cases, $R_{g\perp}$ is larger than $R_{g\parallel}$. This demonstrates the orientation occurs in the direction of imbibition. The capillary filling of polymer chains in nanopores is like pulling a thread in a string ball. This leads to a result that most of the chains run parallel through the capillary in an approximately straight state, as has been observed in semi-dilute solutions[50]. Such chain orientation effects have recently being discussed for the segmental and chain process of cis-1,4-polyisoprene confined in nanoporous aluminum.[51] The combination of the chain orientation and the decrease in average number of entanglement points $Z$ per chain provide further evidence in support of the theory that chains can only move along the "reptation tube" under confinement.

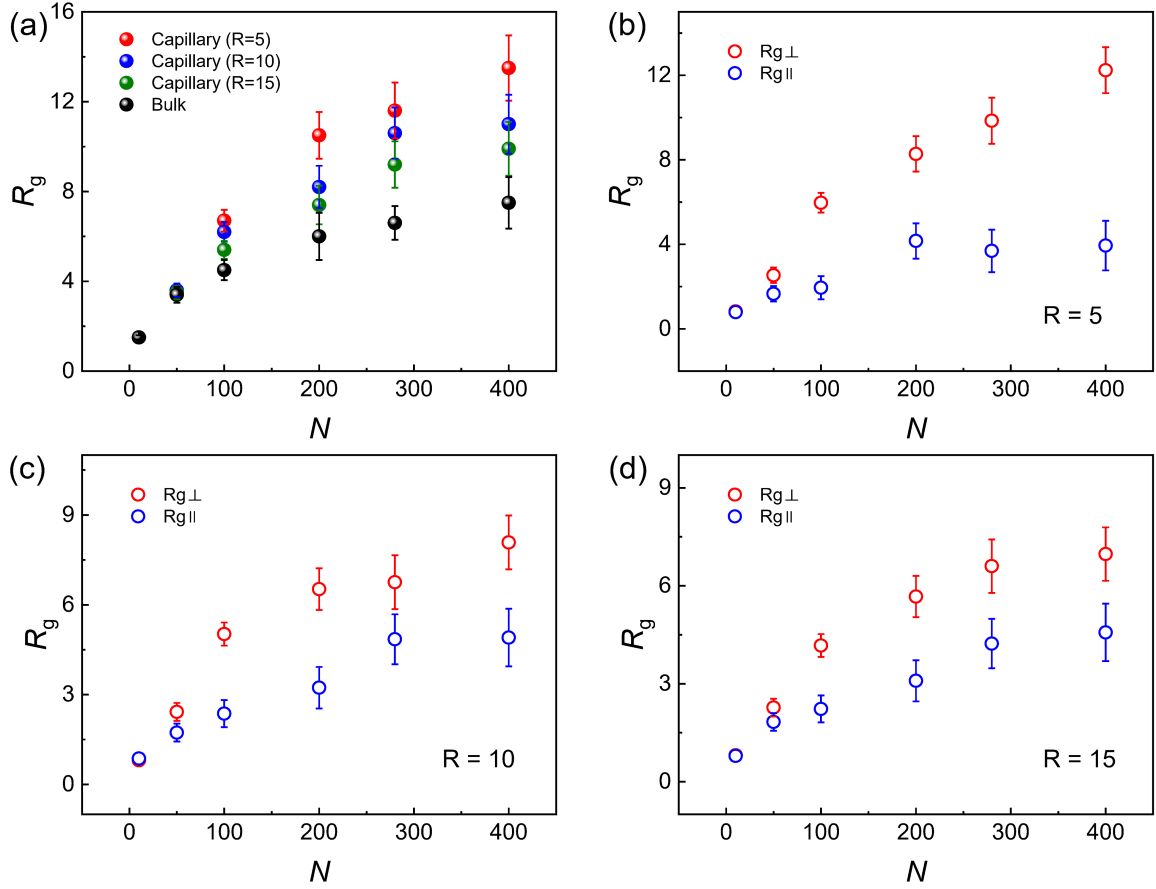

**Figure 11.** (a) Calculated $R_g$ of polymer chains in the capillary with $R = 5\,\sigma$ (red solid dots), $R = 10\,\sigma$ (blue solid dots), $R = 15\,\sigma$ (green solid dots), and in the bulk (black solid dots) respectively. (b) (c) (d) Calculation results of $R_g$ of polymer chains in filling processes in the capillary with (b) $R = 5\,\sigma$, (c) $R = 10\,\sigma$, and (d) $R = 15\,\sigma$. $R_{g\perp}$ (red hollow dots) represents $R_g$ of polymer chains in the direction perpendicular to the cross section of the capillary. $R_{g\parallel}$ (blue hollow dots) represents $R_g$ of polymer chains in the direction parallel to the cross section of the capillary.

Considering all effects together, we calculate the effective viscosity $\eta_{eff}$ from the LWE as $\eta_{eff} = \gamma R\cos\theta / 2A_a^2$. In previous works by Yao *et al.*[20] and Venkatesh *et al.*[15], the scaling was close to $\eta_{eff} \propto N^{1.0}$. In our simulation, by taking the logarithm of $\eta_{eff}$ and $N$ (when $N \geq 10$) and further applying a linear fit, we find a changing slope when nanopore radius. As shown in Fig. 12(a), we obtain the scaling $\eta_{eff} \propto N^{0.33}$ for nanopore radius $R = 5\,\sigma$, $\eta_{eff} \propto N^{0.49}$ for $R = 10\,\sigma$, and $\eta_{eff} \propto N^{0.56}$ for $R = 15\,\sigma$. To better understand these results, we apply the relation about $\eta_{eff}/\eta$ under the nanopore confinement,[15,20]

$$\frac{\eta_{eff}}{\eta} = \left[\left(\frac{R_{eff}}{R}\right)^4 + \frac{\varphi\eta}{NR^2}\right]^{-1} \qquad (10)$$

where $\varphi$ is a constant that contains the bulk polymer properties including the entanglement length. From Eq. (10), is evident that the scaling $\eta_{eff} \propto N^{1.0}$ is valid only when the first term is negligibly small. As $R_{eff}/R$ is also a function of $N$, the exponent in the scaling of $\eta_{eff}$ becomes smaller than 1.0. Furthermore, in the works by Yao *et al.* and Venkatesh *et al.*, the systems were entangled melts with $\eta \propto N^{3.4}$, instead of $\eta \propto N^{1.0}$ in the present case where entanglement effects only start when $N > 200$. This effect can reduce the exponent in the scaling of $\eta_{eff}$. We further plot $\eta_{eff}/\eta$ as a function of $1/R$ as Yao *et al.* and Venkatesh *et al.* did. As shown in Fig. 12(b), the data have a similar trend with the theory when plotted versus $1/R$. First, because the nanopore radius employed in our simulation is relatively small, the curves in Fig. 12(b) exhibit higher effective viscosity than in bulk. This is consistent with the results shown in Fig. 3 of Ref. 20. Additionally, with increasing $N$, the reduction in the viscosity becomes moderate. This illustrates that the value of $1/R$ where the effective viscosity attains its maximum is shifting towards zero."

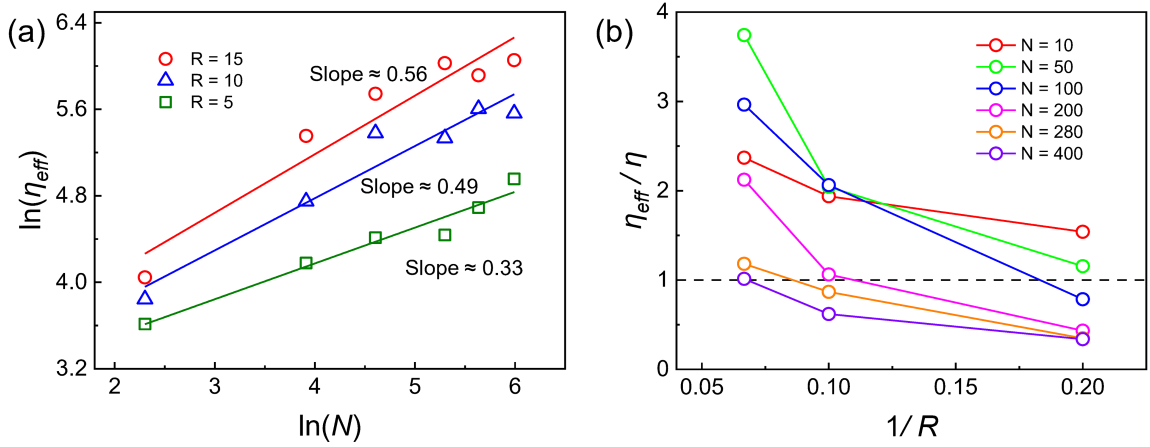

**Figure 12.** (a) Variation of the effective viscosity $\eta_{eff}$ with the degree of polymerization $N$. The slope of linear fitting is about 0.56 for melts in the capillary with $R = 15\,\sigma$ (red line), 0.49 for melts in the capillary with $R = 10\,\sigma$ (blue line), and 0.33 for melts in the capillary with $R = 5\,\sigma$ (green line). (b) Variation of the ratio of effective viscosity $\eta_{eff}$ and bulk viscosity $\eta$, with the reciprocal of the capillary radius $1/R$.

## CONCLUSIONS

In summary, we studied the chain-length effects on the capillary filling dynamics of polymer chains in nanopores. Our model gives a result consistent with the experiments, that is, shorter chains show slower imbibition and longer chains show faster imbibition than the theoretical prediction based on the LWE. The reversal in dynamics is influenced by the change in the number of entanglement points of polymer chains during capillary filling. For short chains with few or no entanglements ($N \leq 50$ in our simulation), the chains are only affected by the slow-filling effects, such as the reduction of the effective radius resulted by viscous areas and the lower free energy caused by confinement. For entangled chains ($N \geq 100$ in our simulation), we find a decrease in $Z$ during capillary filling. This inevitably reduces the viscosity and gives rise to faster filling. This effect is more pronounced in the capillary with the smaller radius. As chain length increases, the acceleration effect reaches a degree comparable to the deceleration effect, and a reversal in the imbibition dynamics occurs as experimentally observed. The increase in $R_g$ of polymer chains and the concomitant chain orientation effects during capillary filling is the direct cause of the lower $Z$. In addition, $R_{g\perp}$ is larger than $R_{g\parallel}$ suggesting that the orientation direction of chains is along the direction of imbibition. Considering the striking effects of the chain length, we calculate the effective viscosity $\eta_{eff}$ to sum up all the influences. Compared with previous works, we find the scaling of bulk viscosity $\eta$, or the degree of entanglement in systems, will influence the scaling of $\eta_{eff}$. We also find that the exponent in the scaling relation between $\eta_{eff}$ and $N$ may increase with increasing capillary radius, as the scaling $N^{1.0}$ is recovered under conditions that the polymer chains are well entangled.

**ASSOCIATED CONTENT**

The following files are available free of charge.

Properties of polymer melt required by LWE, the results of mapping our simulation to a physical system, method of primitive path analysis, and fits of the radius of gyration to the scaling relation (PDF).

Videos of imbibition to verify the brush-like chains and the viscous area near the wall (MP4).

**AUTHOR INFORMATION**

**Corresponding Author**

*Jiajia Zhou - South China Advanced Institute for Soft Matter Science and Technology, School of Emergent Soft Matter, South China University of Technology, Guangzhou 510640, China;

Guangdong Provincial Key Laboratory of Functional and Intelligent Hybrid Materials and Devices, South China University of Technology, Guangzhou 510640, China; Email: zhouj2@scut.edu.cn.

**Authors**

Jianwei Zhang - Faculty of Materials Science and Engineering, South China University of Technology, Guangzhou 510640, China

Jinyu Lei - Faculty of Materials Science and Engineering, South China University of Technology, Guangzhou 510640, China

Wenzhang Tian - Faculty of Materials Science and Engineering, South China University of Technology, Guangzhou 510640, China

Guangzhao Zhang - Faculty of Materials Science and Engineering, South China University of Technology, Guangzhou 510640, China

George Floudas - Max Planck Institute for Polymer Research, 55128 Mainz, Germany; Department of Physics, University of Ioannina, 45110 Ioannina, Greece; Institute of Materials Science and Computing, University Research Center of Ioannina (URCI), 45110 Ioannina, Greece

**Notes**

The authors declare no competing financial interest.

**ACKNOWLEDGMENT**

This research was supported to J.Z. by National Key R&D Program of China (2022YFE0103800), the National Natural Science Foundation of China (21774004), the Recruitment Program of Guangdong (2016ZT06C322) and the 111 Project (B18023). GF acknowledges the Hellenic Foundation for Research and Innovation (H.F.R.I.) under the "First Call for H.F.R.I. Research Projects to support Faculty members and Researchers and the procurement of high-cost research equipment grant" (Project Number: 183).

**Support Information**

# Capillary Filling of Polymer Chains in Nanopores

Jianwei Zhang[1], Jinyu Lei[1], Wenzhang Tian[1], Guangzhao Zhang[1], George Floudas[2,3,4],

Jiajia Zhou[5,6,*]

1. Faculty of Materials Science and Engineering, South China University of Technology, Guangzhou 510640, China
2. Max Planck Institute for Polymer Research, 55128 Mainz, Germany
3. Department of Physics, University of Ioannina, 45110 Ioannina, Greece
4. Institute of Materials Science and Computing, University Research Center of Ioannina (URCI), 45110 Ioannina, Greece
5. South China Advanced Institute for Soft Matter Science and Technology, School of Emergent Soft Matter, South China University of Technology, Guangzhou 510640, China
6. Guangdong Provincial Key Laboratory of Functional and Intelligent Hybrid Materials and Devices, South China University of Technology, Guangzhou 510640, China

**Contents**

1. Surface tension
2. Density profiles
3. Poiseuille flow
4. Viscosity
5. Contact angle
6. Mapping to the physical system
7. Entanglement
8. Radius of gyration in bulk

# 1. Surface Tension

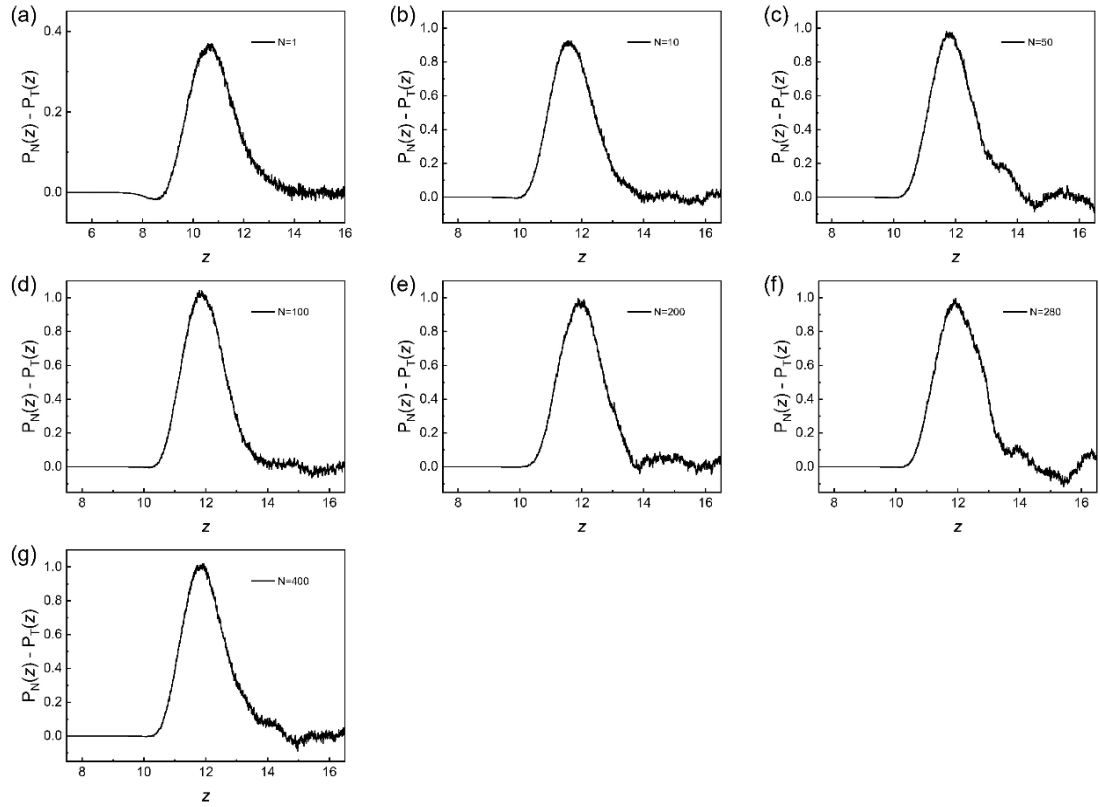

**Figure S1.** $P_N(z) - P_T(z)$ of bottom of model before imbibing in the cases of (a) $N = 1$, (b) $N = 10$, (c) $N = 50$, (d) $N = 100$, (e) $N = 200$, (f) $N = 280$, and (g) $N = 400$. The thickness of whole gas-liquid interface is about $3.5\ \sigma$. The surface tension is calculated by numerical integration.

## 2. Density profiles

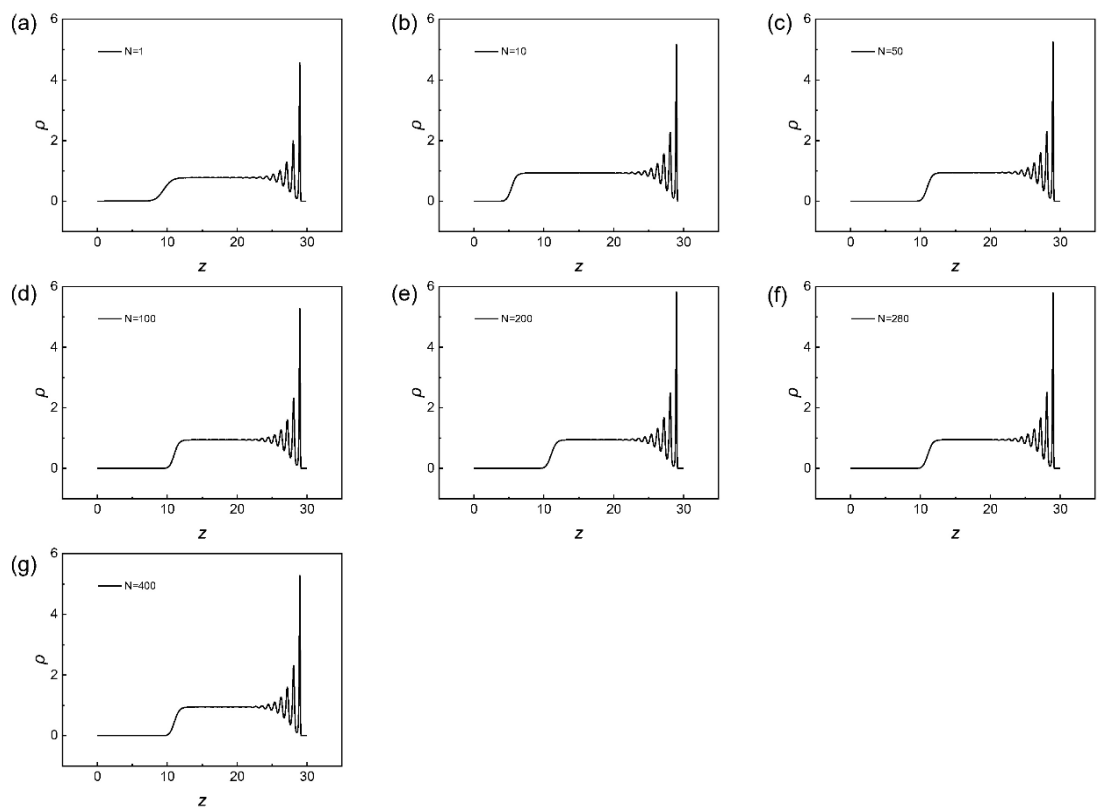

**Figure S2.** Density profiles of bottom of model from $z = 0$ to $z = 30$ before imbibing in the cases of (a) $N = 1$, (b) $N = 10$, (c) $N = 50$, (d) $N = 100$, (e) $N = 200$, (f) $N = 280$, and (g) $N = 400$. It represents density of the vapor, interface, and bulk of polymer melt in turn as $z$ increases. On the far right of figure is the density oscillation of particles approaching the wall.

**Table S1. Density of liquid in bulk phase.**

| N | 1 | 10 | 50 | 100 | 200 | 280 | 400 |
|---|---|---|---|---|---|---|---|
| ρ | 0.777 | 0.930 | 0.942 | 0.944 | 0.946 | 0.947 | 0.949 |

## 3. Poiseuille flow

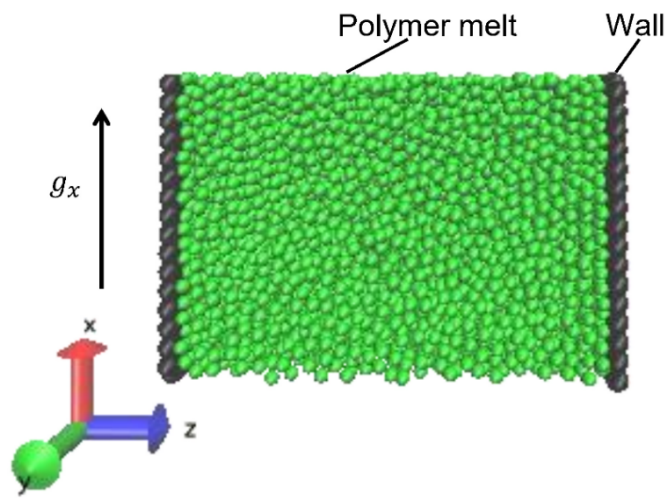

**Figure S3.** Snapshot of Poiseuille flow of polymer melt in the case of $N = 100$.

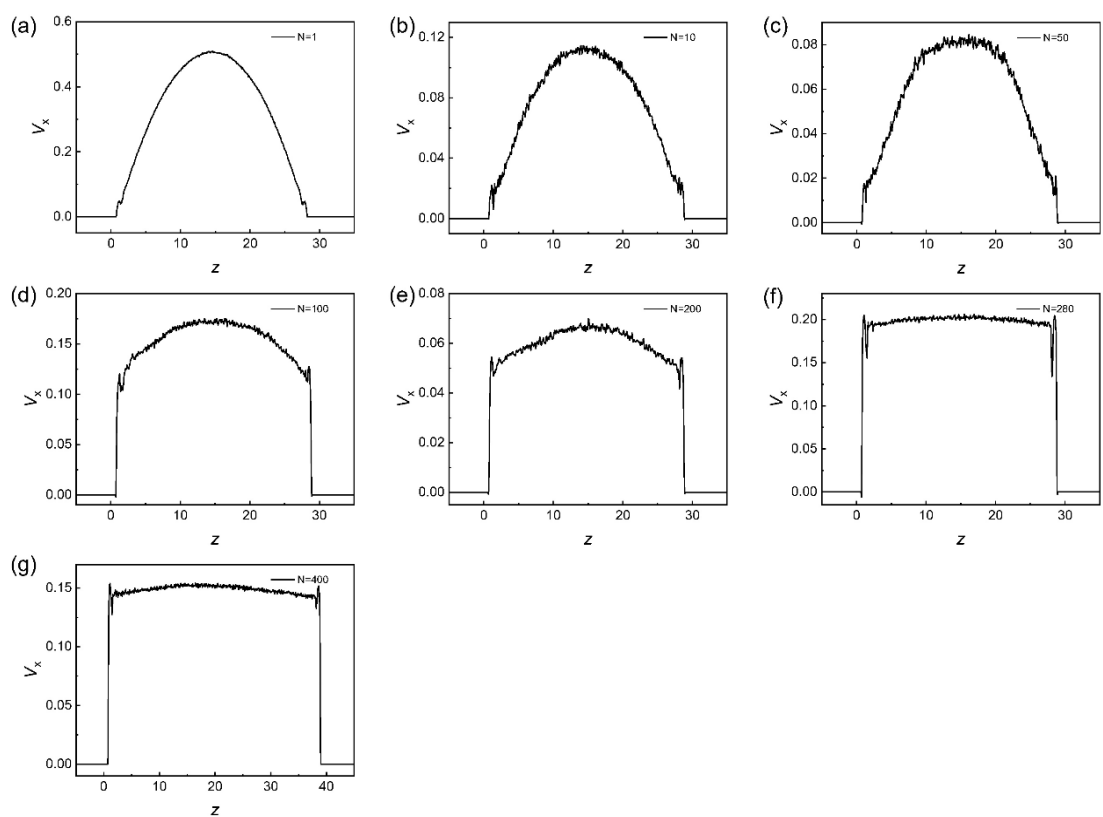

**Figure S4.** Velocity profiles of polymer melt Poiseuille flow along $z$-axis at (a) $N = 1$, (b) $N = 10$, (c) $N = 50$, (d) $N = 100$, (e) $N = 200$, (f) $N = 280$, and (g) $N = 400$. Slip boundary conditions gradually arise as $N$ of polymer increases and only the region in the center that presents a quadratic function is fitted. To make the value of viscosity collecting from velocity profile in Fig. S4 more accurate, a series of different value of $g_x$, from $0.2\ k_B T$ to $0.8\ k_B T$, were used in simulation for polymer melt with different degree of polymerization.

## 4. Viscosity

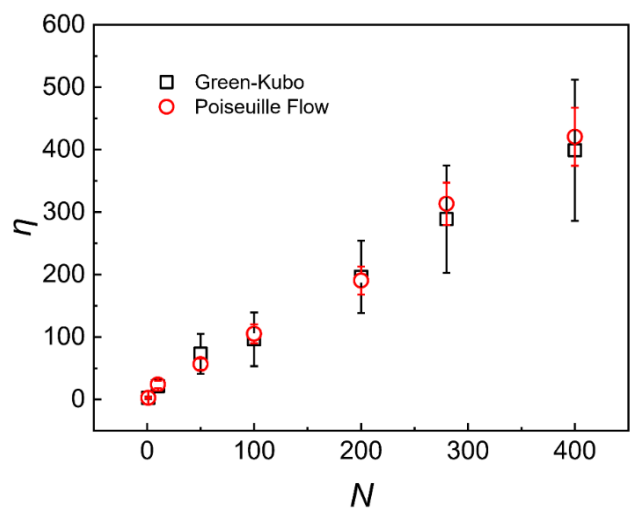

**Figure S5.** The viscosity $\eta$ of polymer melt as a function of the degree of polymerization $N$. The data are calculated by Green-Kubo method (black points) and Poiseuille flow method (red points), respectively.

## 5. Contact Angle

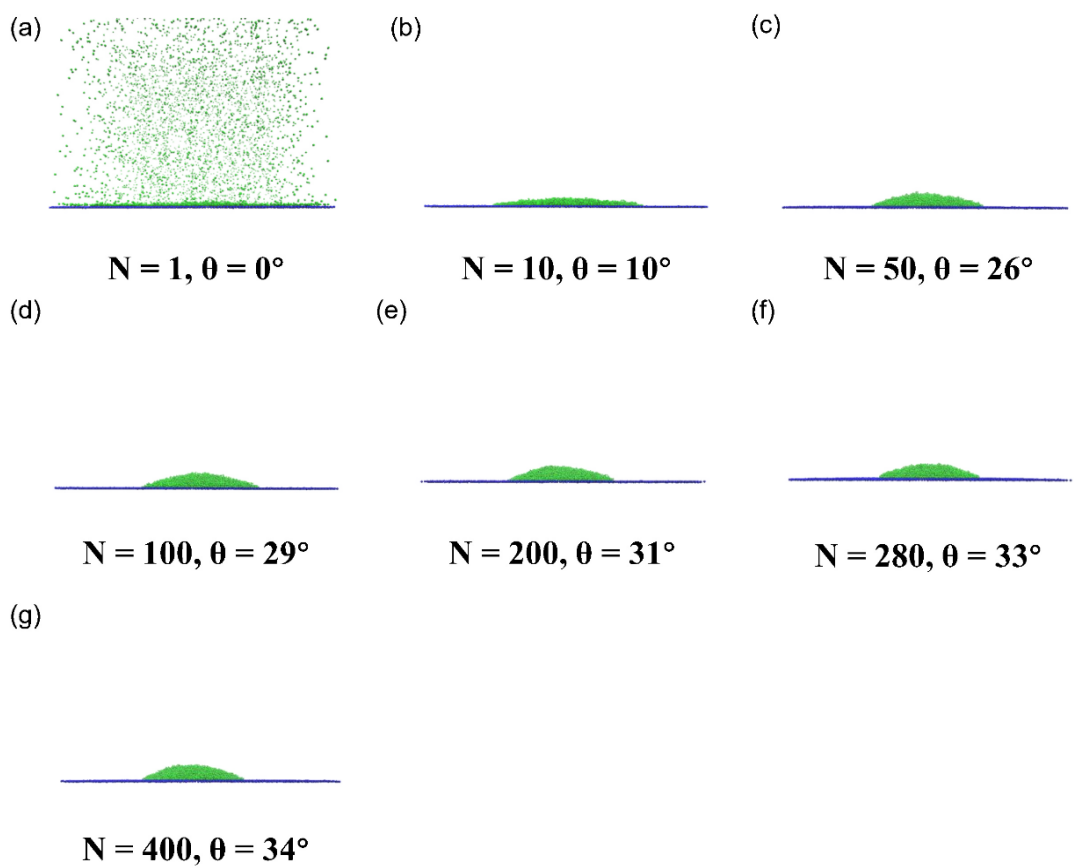

**Figure S6.** Snapshots of polymer droplets after wetting of a plate with same lattice constant as the capillary, at (a) $N = 1$, (b) $N = 10$, (c) $N = 50$, (d) $N = 100$, (e) $N = 200$, (f) $N = 280$, and (g) $N = 400$. The interaction parament between polymer and wall atoms is $\varepsilon_{LJ} = 1.6\ k_B T$.

## 6. Mapping to the physical system

We can attempt to map our simulation system to the real physical system. According to the poly(ethylene oxide) (PEO) system used in the work by Yao et al.[1], the $N = 400$ system in our simulation is roughly corresponding to the $N = 1600$ system in the physical case, thus one simulation bead corresponds 4 EO groups. We evaluate the length scale $4\pi\sigma^3/3 = 4v_0$, where $v_0 = 6.04 \times 10^{-29}$ m is the monomer volume. The length scale is given by $\sigma \approx 3.9$ Å. The unit of mass then can be obtained as $m = 4M_0/N_A = 4.1 \times 10^{-25}$ kg, where $M_0 = 62$ g/mol is the monomer molecular weight and $N_A = 6.02 \times 10^{23}$ mol-1 is the Avogadro's constant. The energy scale is given by $\epsilon = k_B T \approx 5.0 \times 10^{-21}$ J with $T = 358$ K. The time unit is then can be calculated as $\tau = \sqrt{m\sigma^2/\epsilon} = 3.5 \times 10^{-12}$ s. When value of surface tension in the simulation $\gamma_{MD}$ equals to 1.71, value of surface tension in the physical system can be given as $\gamma_{real} = \gamma_{MD}\epsilon/\sigma^2 = 57$ mN/m, which is comparable to the physical system $\gamma = 27.8$ mN/m. We can also calculate the friction parameter to be $\xi = 0.5\, m/\tau = 5.8 \times 10^{-12}$ kg/s. When value of viscosity in the simulation $\eta_{MD}$ equals to 420.6, value of viscosity in the physical system can be given as $\eta_{real} = \eta_{MD}\sqrt{m\epsilon/\sigma^4} = 0.13$ Pa·s. This is significantly smaller than the real value (in the kPa·s range). This indeed suggests a complete understanding of the imbibition dynamics would require further studies for varying friction parameter $\xi$.

## 7. Entanglement

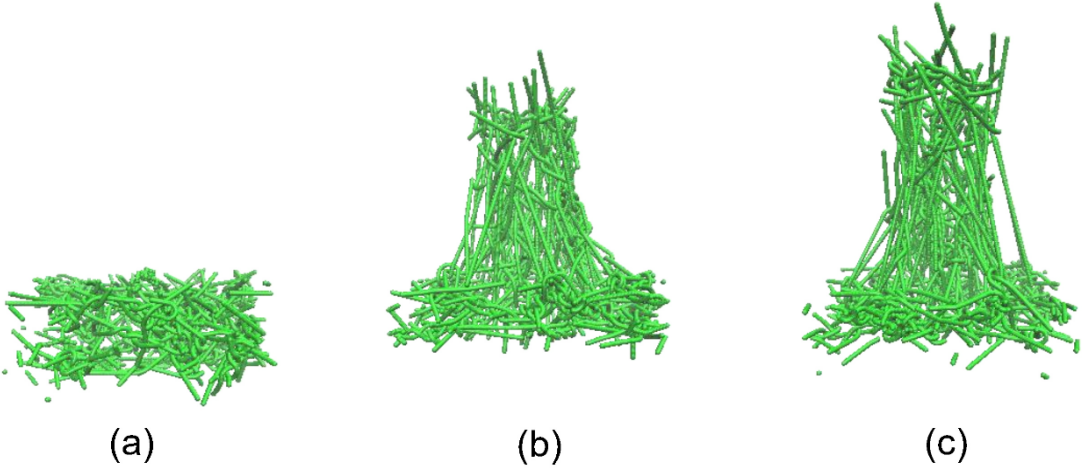

(a)  (b)  (c)

**Figure S7.** Snapshots of calculation processes of the number of entanglement points $Z$ in the case of $N = 200$ when $R = 10\,\sigma$ at (a) $t = 0$; (b) $t = 30000$; (c) $t = 40000$.

The specific operation in simulation is as described by work of Everaers[2]: To begin with, two beads in the end position of every chain are fixed in space. Next, the intrachain excluded volume interactions are canceled, while retaining the interchain excluded-volume interactions. Finally, slowly freeze the system toward $T = 0$ to minimize the energy of the system. Topology files used for each calculation are obtained in capillary filling processes at $t = 0, 25000, 30000, 35000, 40000$ respectively. The entanglement length $N_e$ is then collected by employing

$$N_e = (N - 1)\left(\frac{<L_{pp}^2>}{<R_{pp}^2>} - 1\right)^{-1} \quad (1)$$

where $L_{pp}$ is the contour length and $R_{pp}$ is the mean-square end-to-end distance. The number of entanglement points $Z$ is then calculated by $Z = \frac{N}{N_e}$.

## 8. Radius of gyration in bulk

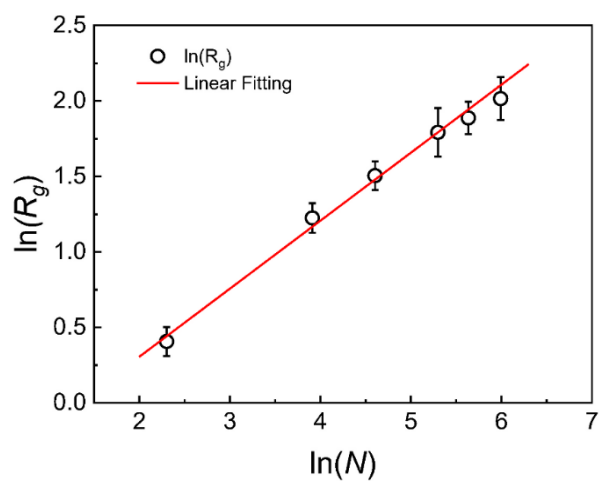

**Figure S8.** Variation of the radius of gyration $R_g$ of polymers in bulk with degree of polymerization $N$. The red line is the result of linear fitting, the slope is about 0.5.